\documentclass[structabstract]{aa}
\usepackage{graphicx}
\usepackage{txfonts}
\usepackage{natbib}
\usepackage{ulem}

\bibpunct[]{(}{)}{;}{a}{}{,}

\begin{document}

   \title{Laboratory spectroscopic study of the $^{15}$N isotopomers of cyanamide, 
          H$_2$NCN, and a search for them toward IRAS 16293$-$2422~B\thanks{The 
          experimental line lists are available at the CDS via anonymous 
          ftp to cdsarc.u-strasbg.fr (130.79.128.5) or via 
          http://cdsweb.u-strasbg.fr/cgi-bin/qcat?J/A+A/623/A93}}

   \author{Audrey Coutens\inst{1}
           \and
           Olena Zakharenko\inst{2}
           \and
           Frank Lewen\inst{2}
           \and
           Jes K. J{\o}rgensen\inst{3}
           \and
           Stephan Schlemmer\inst{2}
           \and
           Holger S.~P. M{\"u}ller\inst{2}
           }

   \institute{Laboratoire d’Astrophysique de Bordeaux, Univ. Bordeaux, CNRS, 
              B18N, all{\'e}e Geoffroy Saint-Hilaire, 33615 Pessac, France\\
              \email{audrey.coutens@u-bordeaux.fr} 
              \and
              I.~Physikalisches Institut, Universit{\"a}t zu K{\"o}ln,
              Z{\"u}lpicher Str. 77, 50937 K{\"o}ln, Germany\\
              \email{hspm@ph1.uni-koeln.de} 
              \and
              Centre for Star and Planet Formation, Niels Bohr Institute and Natural History Museum 
              of Denmark, University of Copenhagen, {\O}ster Voldgade 5$-$7, 1350 Copenhagen K, Denmark
              }

   \date{Received 08 November 2018 / Accepted 20 December 2018}
 
  \abstract
{Cyanamide is one of the few interstellar molecules containing two chemically different N atoms. 
It was detected recently toward the solar-type protostar IRAS 16293$-$2422~B together with 
H$_2$N$^{13}$CN and HDNCN in the course of the Atacama Large Millemeter/submillimeter Array 
(ALMA) Protostellar Interferometric Line Survey (PILS). The detection of the $^{15}$N 
isotopomers or the determination of upper limits to their column densities was hampered by 
the lack of accurate laboratory data at the frequencies of the survey.}
{We wanted to determine spectroscopic parameters of the $^{15}$N isotopomers of cyanamide 
that are accurate enough for predictions well into the submillimeter region and to search 
for them in the PILS data.}
{We investigated the laboratory rotational spectra of H$_2^{15}$NCN and H$_2$NC$^{15}$N 
in the selected region between 192 and 507~GHz employing a cyanamide sample in natural 
isotopic composition.  Additionally, we recorded transitions of H$_2$N$^{13}$CN.}
{We obtained new or improved spectroscopic parameters for the three isotopic species. Neither 
of the $^{15}$N isotopomers of cyanamide were detected unambiguously in the PILS data. Two 
relatively clean lines can be tentatively assigned to H$_2^{15}$NCN. If confirmed, their 
column densities would imply a low $^{14}$N/$^{15}$N ratio for cyanamide toward this source.}
{The resulting line lists should be accurate enough for observations up to about 1~THz. 
More sensitive observations, potentially at different frequencies, may eventually lead 
to the astronomical detection of these isotopic species.}

\authorrunning{A. Coutens et al.}
\titlerunning{Laboratory spectroscopic study of isotopic H$_2$NCN}

\maketitle
\hyphenation{For-schungs-ge-mein-schaft}
\hyphenation{Schott-ky}
\hyphenation{cyan-a-mide}

\section{Introduction}
\label{intro}

Cyanamide, H$_2$NCN, is one of the few interstellar molecules that contain two chemically 
inequivalent N atoms.\footnote{The other are N$_2$H$^+$, N$_2$O, aminoacetonitrile (NH$_2$CH$_2$CN), 
a heavy homolog of cyanamide, and $E$-cyanomethanimine ($E$-HNCHCN), see e.g., the Molecules 
in Space page (https://cdms.astro.uni-koeln.de/classic/molecules) of the Cologne Database for Molecular 
Spectroscopy \citep{CDMS_3}, or the list of Interstellar and Circumstellar Molecules of the 
Astrochymist page (http://www.astrochymist.org/astrochymist\_ism.html).} It was among the molecules 
detected early in the interstellar medium (ISM) by radio astronomical means \citep{H2NCN_det_1975}. 
The detection was made toward the prolific high-mass star-forming region Sagittarius (Sgr) B2, 
which is close to the Galactic center. There is also a report on the detection of H$_2$NCN toward 
the nearby Orion-KL region \citep{H2NCN_Orion-KL_2003} and the high-mass protostar IRAS 20126+410 
\citep{Palau2017}. Cyanamide was also found in two nearby starburst galaxies NGC~253 
\citep{H2NCN_extra-gal_2006} and M82 \citep{H2NCN_M82_2011}.

\citet{H2NCN_etc_solar-type_2018} reported the first detection of H$_2$NCN toward two solar-type 
protostars; one was IRAS 16293$-$2422~B with the Atacama Large Millimeter/submillimeter Array 
(ALMA) in the framework of the Protostellar Interferometric Line Survey (PILS), and the other 
was NGC~1333 IRAS2A in the course of Plateau de Bure Interferometer (PdBI) observations. 
More recently, a tentative detection was also reported toward the low-mass protostellar source 
Barnard~1b \citep{Marcelino2018}.

Isotopic fractionation in the N isotopes has attracted considerable interest in recent years. 
Very different $^{14}$N/$^{15}$N ratios have been determined in star-forming regions which range 
from less than 100 to around 1000 \citep[e.g.,][]{Milam2012,Bizzocchi2013,Wampfler2014}, compared 
to the terrestrial ratio of 272 \citep{iso-comp_2011}. However, the reasons for the enrichment or 
depletion in $^{15}$N are not clear at present.

In order to address the possible N-fractionation for cyanamide, reliable transition frequencies 
of these rarer isotopologs are critical. The laboratory rotational spectrum of the H$_2$NCN main 
isotopolog has been studied extensively since the fairly early days of microwave spectroscopy 
\citep{H2NCN_rot_1959}, especially by \citet{H2NCN_rot_1986} and by \citet{H2NCN_FIR_1993}. 
Only very limited information was available for minor isotopologs of cyanamide 
\citep{H2NCN_isos_rot_1972,H2NCN_isos_rot_1985} until fairly recently when 
\citet{H2NCN_isos_rot_2011} reported millimeter and in part submillimeter transition 
frequencies for numerous isotopic species. \citet{H2_HD_D2NCN_rot_2013} presented updated 
accounts on H$_2$NCN, HDNCN, and D$_2$NCN. The line lists of the $^{15}$N isotopomers covered 
135$-$176~GHz ($J'' = 6$ to 8) besides some older, lower-accuracy data with $J'' = 0$ or 
1 \citep{H2NCN_isos_rot_2011}. Therefore, we decided to record rotational transitions of 
H$_2^{15}$NCN and H$_2$NC$^{15}$N between 192 and 507~GHz which should be sufficient to 
search for these isotopic species throughout the entire presently available frequency range 
of ALMA up to almost 1~THz. In the course of our measurements, we also recorded some 
transitions of H$_2$N$^{13}$CN to improve its spectroscopic parameters somewhat.


\begin{figure}
\centering
 \includegraphics[width=8cm,angle=0]{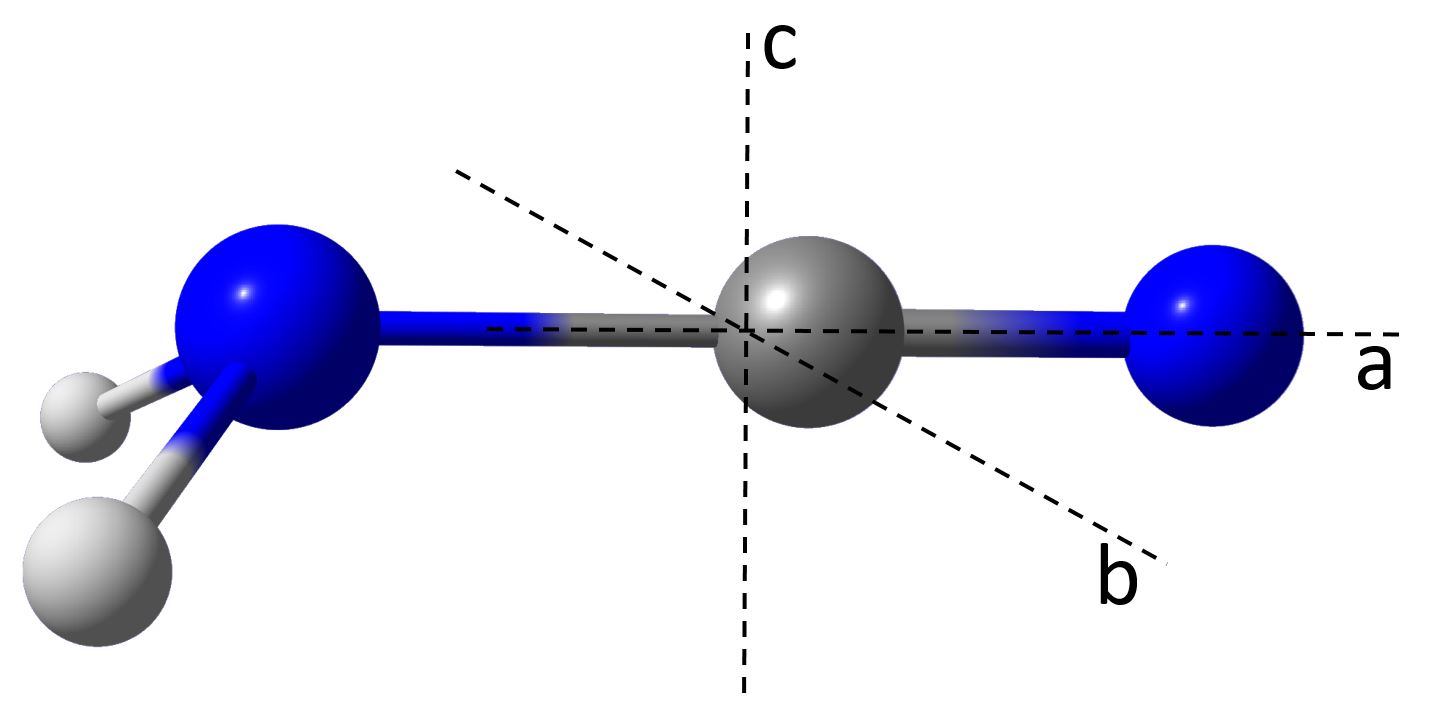}

\caption{
   Schematic representation of the cyanamide molecule; C and N atoms are indicated by gray 
   and blue spheres, respectively, and H atoms by smaller, light gray spheres. The approximate 
   positions of the inertial axes are represented by dashed lines. We note that the C atom is 
   very close to the center of mass and the N atoms are very close to the $a$-axis.}
\label{cyanamide_structure}
\end{figure}

\section{Spectroscopic properties of cyanamide}
\label{rot_vib_backgr}

Cyanamide is an asymmetric rotor with $\kappa = (2B - A - C)/(A - C) = -0.9983$, very close to 
the symmetric limit of $-1$. A model of the molecule is shown in Fig~\ref{cyanamide_structure}. 
It is nonplanar with two equivalent configurations, albeit with a low barrier to planarity 
\citep{H2_HD_D2NCN_rot_2013}. Tunneling between the 
two minima lifts the degeneracy and creates tunneling substates $0^+$ and a $0^-$ which 
are separated by 1486~GHz (or 49.57~cm$^{-1}$ or 71.32~K). 
The symmetry of the planar transition state is $C_{\rm 2v}$ with the $a$-axis being the 
symmetry axis. The spin-statistics of the two equivalent H nuclei lead to \textit{ortho} 
and \textit{para} states with a spin-statistical weight ratio of 3\,:\,1. 
The \textit{ortho} states are described by $K_a$ being even in $0^+$ and odd in $0^-$. 
Transitions within the tunneling states obey $a$-type selection rules with 
$\mu_a(0^+) = (4.25 \pm 0.02)$~D and $\mu_a(0^-) = (4.24 \pm 0.02)$~D 
\citep{H2NCN_isos_rot_1985}. Tunneling occurs along the $c$-inertial axis, hence the two 
tunneling states are connected by $c$-type transitions with $\mu_c = (0.91 \pm 0.02)$~D 
\citep{H2NCN_isos_rot_1985}.

Carbon has two stable isotopes with mass numbers 12 and 13; the terrestrial abundances of 
the latter is 1.11\,\% \citep{iso-comp_2011}. The corresponding abundance of $^{15}$N is 
0.36\,\%, much less than the dominant $^{14}$N.


\begin{figure}
\centering
 \includegraphics[width=8cm,angle=0]{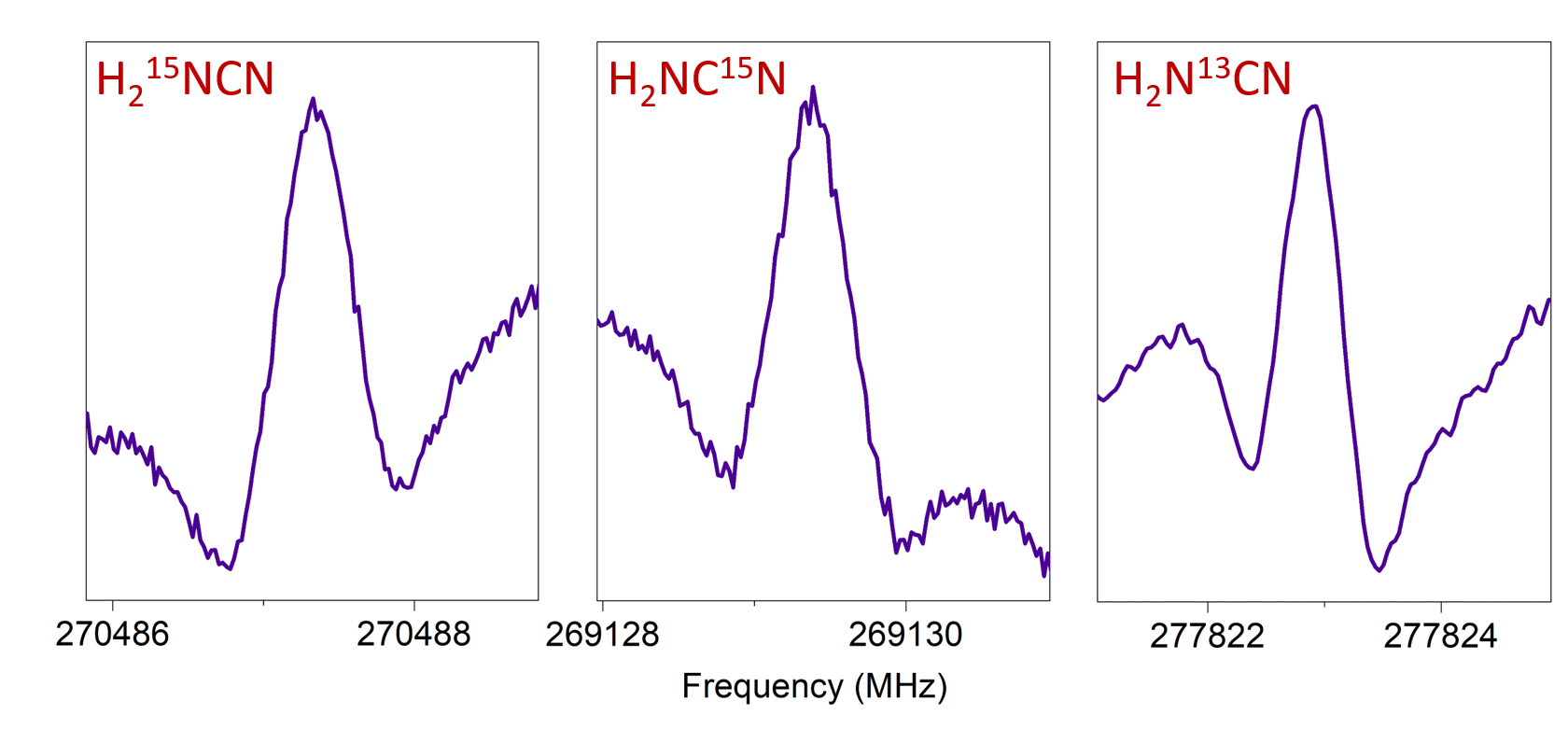}

\caption{Sections of the rotational spectrum of cyanamide displaying the 
   $J_{K_a,K_c} = 14_{1,14} - 13_{1,13}$ transitions of three isotopic species within 
   the $0^-$ tunneling state.}
\label{lab-spec}
\end{figure}

\section{Laboratory spectroscopic details and observed lines}
\label{exptl}

The investigations of isotopic cyanamide were carried out in a five-meter long double-path glass 
cell at room temperature and under slow flow conditions. A commercial sample from 
Sigma-Aldrich was used without further purification. The sample required slight heating 
to reach a sufficient vapor pressure which was maintained at about 0.5~Pa inside the cell. 
We employed three frequency multipliers (Virginia Diodes, Inc.) driven by an Agilent 
E8257D synthesizer as sources with appropriate Schottky diodes as detectors to cover 
frequencies between 192 and 507~GHz. Frequency modulation reduces standing waves, 
and the demodulation at twice the modulation frequency leads to approximately 
second-derivative line-shapes. Further details on the spectrometer system are available 
in \citet{OSSO_Spektrm_2015}.

We optimized the conditions, measurement time, pressure, and modulation deviation, by 
recording selected transitions of H$_2$N$^{13}$CN with $K_a$ values of 1 and 2 around 
300~GHz which were well predicted by the available data \citep{H2NCN_isos_rot_2011}. 
An integration time of 60~ms per measurement point of one scan upward in frequency and 
one downward was sufficient for these lines and most of the $^{15}$N isotopomers. 
Several scans were coadded for some weaker lines.

The available H$_2^{15}$NCN and H$_2$NC$^{15}$N transition frequencies obeyed $a$-type 
selection rules ($\Delta K_a = 0$ and $\Delta K_c = 1$). They covered $J = 7 - 6$ to $9 - 8$ 
with $K_a \le 5$ and 4 for 0$^+$ and 0$^-$, respectively, from \citet{H2NCN_isos_rot_2011}, 
and some $J = 1 - 0$ and $2 - 1$ transitions from \citet{H2NCN_isos_rot_1972} and 
\citet{H2NCN_isos_rot_1985}. The resulting spectroscopic parameters were good enough to 
find transitions with low values of $J$, starting with $10 - 9$ around 195~GHz and low 
values of $K_a$. We determined new spectroscopic parameters and covered several higher 
$J$ up to $26 - 25$ near 505~GHz. The $K_a$ range extended eventually to 7 and 6 for 
0$^+$ and 0$^-$, respectively. The uncertainties ranged from 20~kHz for isolated lines 
with symmetric line shape and good signal-to-noise ratio to 100~kHz.

Subsequently, we recorded H$_2$N$^{13}$CN transitions to supplement earlier measurements 
by \citet{H2NCN_isos_rot_2011}. Their data included $J = 6 - 5$ to $8 - 7$ and $29 - 28$ 
to $32 - 31$ with $K_a \le 5$ and 6 for 0$^+$ and 0$^-$, respectively, at frequencies of 
119 to 161 and 578 to 641~GHz. Our data covered $J = 13 - 12$ to $25 - 24$ and mostly 
higher $K_a$ values, 6 for 0$^-$ and 7 for 0$^+$ and 0$^-$, along with some lower-$K_a$ data. 
We assigned uncertainties of 20 or 30~kHz to our H$_2$N$^{13}$CN transition frequencies. 
Figure~\ref{lab-spec} displays the $0^-$, $J_{K_a,K_c} = 14_{1,14} - 13_{1,13}$ transition 
of each of the three isotopologs.

\section{Determination of spectroscopic parameters}
\label{lab-results}

We used Pickett's SPCAT and SPFIT programs \citep{spfit_1991} to predict and fit rotational 
spectra of the minor isotopic species of cyanamide employing the reduced axis system 
Hamiltonian \citep{RAS_1972}, as was done previously 
\citep{H2NCN_rot_1986,H2NCN_FIR_1993,H2NCN_isos_rot_2011,H2_HD_D2NCN_rot_2013}. 
We used Watson's $S$ reduction of the rotational Hamiltonian, in contrast to the latest 
two publications but in agreement with earlier work. 
Noting that the ground vibrational state $\varv = 0$ consists of the substates 0$^+$ and 
0$^-$, we fit average spectroscopic parameters $X = (X(0^+) + X(0^-))/2$ and 
$\Delta X = X(0^-) - X(0^+)$, as proposed by \citet{aGg_RAS-formulation_2003}. 
The model was tested for the rotational spectrum of NH$_2$D and applied successfully in an 
analysis of the rotational and rovibrational spectra of H$_2$DO$^+$ \citep{H2DO+_analysis_2010}. 
The advantage of being able to add parameters or their differences separately was 
particularly apparent in the case of ethanethiol \citep{ROH_RSH_2016}.
As the available experimental lines of the minor isotopic species are considerably smaller 
than those of the main species, we estimated several of the higher-order parameters of 
the former by scaling parameters of the main isotopolog by the ratios of appropriate 
powers of $A - (B+C)/2$, $B + C$, and $B - C$, as done recently for H$_2$C$^{16}$O, 
H$_2$C$^{17}$O, and H$_2$C$^{18}$O \citep{H2CO-X_rot_2017}. A similar approach was taken 
for fitting spectra of isotopic species of cyanamide \citep{H2NCN_isos_rot_2011}.


\begin{table*}
\begin{center}
\caption{Spectroscopic parameters $X^a$ (MHz) and their differences $\Delta X^b$ (MHz) of cyanamide isotopologs.}
\label{tab-parameters_CA}
\renewcommand{\arraystretch}{1.20}
\begin{tabular}[t]{lr@{}lr@{}lr@{}lr@{}l}
\hline \hline
Parameter & \multicolumn{2}{c}{H$_2^{15}$NCN} & \multicolumn{2}{c}{H$_2$NC$^{15}$N} & \multicolumn{2}{c}{H$_2$N$^{13}$CN} & \multicolumn{2}{c}{H$_2$NCN} \\
\hline
$\Delta E$                     & 1460174&.~(120)      & 1486116&.~(118)      & 1486522&.~(66)       & 1486004&.275~(185)    \\
$F_{ac}$                       &     342&.505~(46)    &     335&.828~(47)    &     345&.963~(59)    &     346&.5304~(245)   \\
$F_{ac}^K$                     &    $-$1&.442         &    $-$1&.415         &    $-$1&.457         &    $-$1&.4577~(40)    \\
$F_{ac}^J \times 10^3$         &       1&.2756~(113)  &       1&.1048~(119)  &       1&.2514~(108)  &       1&.2009~(89)    \\
$F_{ac}^{JK} \times 10^6$      &   $-$48&.69          &   $-$47&.52          &   $-$50&.54          &   $-$50&.57~(155)     \\
$F_{ac}^{JJ} \times 10^9$      &      10&.64          &      10&.32          &      11&.34          &      11&.36~(48)      \\
$A$                            &  307819&.7~(241)     &  308328&.7~(239)     &  308205&.1~(122)     &  308298&.054~(66)     \\
$\Delta A/2$                   & $-$3838&.00          & $-$3844&.35          & $-$3842&.81          & $-$3843&.971~(33)     \\
$B$                            &    9845&.11174~(75)  &    9794&.54528~(82)  &   10117&.07868~(93)  &   10121&.205258~(188) \\
$\Delta B/2$                   &    $-$8&.74933~(71)  &    $-$8&.09439~(79)  &    $-$8&.56394~(93)  &    $-$8&.551540~(113) \\
$C$                            &    9604&.18856~(76)  &    9555&.59774~(79)  &    9862&.34933~(114) &    9866&.291773~(186) \\
$\Delta C/2$                   &    $-$0&.28904~(77)  &    $-$0&.26534~(81)  &    $-$0&.35051~(110) &    $-$0&.368180~(109) \\
$D_K$                          &      36&.036         &      36&.110         &      36&.070         &      36&.0421~(44)    \\
$\Delta D_K/2$                 &    $-$8&.084         &    $-$8&.101         &    $-$8&.092         &    $-$8&.0857~(33)    \\
$D_{JK} \times 10^3$           &     368&.428~(80)    &     357&.939(~80)    &     377&.325~(101)   &     377&.705~(40)     \\
$\Delta D_{JK}/2 \times 10^3$  &   $-$19&.239~(44)    &   $-$17&.766~(42)    &   $-$18&.097~(83)    &   $-$18&.494~(26)     \\
$D_J \times 10^3$              &       3&.59457~(39)  &       3&.50397~(38)  &       3&.75968~(31)  &       3&.759689~(180) \\
$\Delta D_J/2 \times 10^3$     &       0&.01490~(34)  &       0&.01741~(32)  &       0&.01813~(26)  &       0&.018511~(49)  \\
$d_1 \times 10^6$              &  $-$126&.84~(48)     &  $-$117&.29~(49)     &  $-$129&.70~(36)     &  $-$130&.2699~(176)   \\
$\Delta d_1/2 \times 10^6$     &       8&.89~(47)     &       9&.37~(49)     &      10&.88~(36)     &      10&.5633~(225)   \\
$d_2 \times 10^6$              &   $-$25&.100~(283)   &   $-$23&.484~(304)   &   $-$27&.739~(219)   &   $-$27&.2965~(45)    \\
$\Delta d_2/2 \times 10^6$     &       4&.931~(150)   &       4&.062~(206)   &       5&.516~(165)   &       5&.2987~(43)    \\
$H_K \times 10^3$              &      10&.76          &      10&.79          &      10&.77          &      10&.760~(84)     \\
$\Delta H_K/2 \times 10^3$     &    $-$6&.45          &    $-$6&.47          &    $-$6&.46          &    $-$6&.453~(78)     \\
$H_{KJ} \times 10^6$           &  $-$219&.2~(50)      &  $-$267&.7~(48)      &  $-$196&.2~(45)      &  $-$264&.41~(234)     \\
$\Delta H_{KJ}/2 \times 10^6$  &      71&.6~(12)      &      70&.6~(10)      &      82&.3~(7)       &      73&.64~(180)     \\
$H_{JK} \times 10^6$           &       0&.889~(73)    &       1&.115~(70)    &       0&.945~(44)    &       1&.0798~(167)   \\
$\Delta H_{JK}/2 \times 10^6$  &    $-$0&.250~(27)    &    $-$0&.268~(23)    &    $-$0&.297~(11)    &    $-$0&.2879~(65)    \\
$H_J \times 10^9$              &    $-$0&.761         &    $-$0&.750         &    $-$0&.825         &    $-$0&.826~(77)     \\
$\Delta H_J/2 \times 10^9$     &       0&.189         &       0&.186         &       0&.204         &       0&.205~(9)      \\
$L_K \times 10^6$              &      16&.00          &      16&.07          &      16&.04          &      16&.01           \\
$\Delta L_K/2 \times 10^6$     &    $-$5&.31          &    $-$5&.34          &    $-$5&.32          &    $-$5&.32           \\
$L_{KKJ} \times 10^6$          &    $-$4&.118~(88)    &    $-$3&.439~(82)    &    $-$5&.800~(54)    &    $-$3&.681~(46)     \\
$\Delta L_{KKJ}/2 \times 10^6$ &       0&.343         &       0&.343         &       0&.353         &       0&.353~(39)     \\
$L_{JK} \times 10^9$           &      17&.78~(223)    &      10&.10~(199)    &      17&.44~(94)     &      12&.47~(42)      \\
rms                            &       0&.040         &       0&.037         &       0&.043         &       0&.098          \\
$\sigma ^b$                    &       0&.82          &       0&.86          &       0&.98          &       0&.94           \\
\hline
\end{tabular}
\end{center}
\tablefoot{
Watson's $S$ reduction has been used in the representation $I^r$. 
$^{(a)}$ Numbers in parentheses are one standard deviation in units of the least significant figures. 
         Parameters without uncertainties were estimated and kept fixed in the analyses. 
         $X = (X(0^+) + X(0^-))/2$ and $\Delta X  = X(0^-) - X(0^+)$. 
$^{(b)}$ Reduced standard deviation (unitless).
}
\end{table*}

Our reanalysis of H$_2$NCN data started with the line list from \citet{H2_HD_D2NCN_rot_2013}. 
The very extensive line list was not reproduced within experimental uncertainties on average, 
such that the authors omitted transitions with residuals between observed transition frequencies 
and those calculated from the final set of spectroscopic parameters exceeding three times 
the uncertainties. Initially, we weighted down these lines, but omitted most of them in the end 
because the residuals in our final fit were very similar to those of the previous fit. 
In fact, our final parameter set differed effectively only slightly from that obtained by 
\citet{H2_HD_D2NCN_rot_2013}. 
We added fixed values of $L_K$ and $\Delta L_K$ to account better for transitions 
higher in $K_a$, especially $c$-type transitions between the tunneling states. 
The parameter values were obtained from a fit to the line list deposited in the JPL 
catalog archive \citep{JPL-catalog_1998} which was based on \citet{H2NCN_rot_1986} and 
\citet{H2NCN_FIR_1993} and which extended to transitions considerably higher in $K_a$. 
In addition, we did not use $\Delta L_{JK}$ in the fit. Our parameter values were in good 
agreement with those of \citet{H2_HD_D2NCN_rot_2013}, as was expected. 
The majority of their transition frequencies were obtained by far-infrared spectroscopy. 
They were reproduced on average to $\sim$0.00023~cm$^{-1}$, slightly above the reported 
uncertainties of $\sim$0.00020~cm$^{-1}$. Therefore, we increased the uncertainties 
to $\sim$0.00025~cm$^{-1}$. The parameter values changed little, but now not only 
the frequencies determined with microwave accuracy were reproduced within uncertainties 
on average, but also the far-infrared transition frequencies.

The details for determining spectroscopic parameters were the same for the two $^{15}$N 
isotopomers of cyanamide and for H$_2$N$^{13}$CN. The parameter set of the main isotopic 
species was taken as a start. Low-order spectroscopic parameters were fit until the 
transition frequencies were reproduced within around five times the experimental uncertainties. 
At each step, we searched for the parameter which improved the quality of the fit the most. 
Subsequently, parameters still kept fixed were scaled with appropriate powers of the ratios 
of $B + C$ and $B - C$. Since $A$ was correlated with $\Delta E$, scaling with appropriate 
powers of the ratios of $A - (B + C)/2$ was done subsequently; $\Delta A$ was scaled with 
the ratio of $A$; the distortion corrections to $F_{ac}$ were scaled in addition with 
the ratios of the $F_{ac}$. In the next step, parameters of higher order were released 
until the qualities of the fits no longer significantly improved. 
In a final step, $\Delta A$ was scaled again with the ratio of $A$. 
\citet{H2NCN_isos_rot_2011} assigned uncertainties of 50~kHz to their transition 
frequencies. These were appropriate in the case of the $^{15}$N isotopomers. 
In the case of H$_2$N$^{13}$CN, they were slightly pessimistic at lower frequencies 
(119$-$161~GHz) and slightly optimistic at higher frequencies (578$-$644~GHz). 
Uncertainties of 30 and 60~kHz, respectively, turned out to be appropriate. 
Uncertainties of 100~kHz were used for the few older transition frequencies 
\citep{H2NCN_isos_rot_1972,H2NCN_isos_rot_1985}. The transition frequencies of each 
isotopic species as well as subsets thereof were reproduced mostly within and always 
close to the experimental uncertainties on average. The resulting spectroscopic parameters 
are given in Table~\ref{tab-parameters_CA} together with those of the main isotopic species. 
The experimental line lists have been deposited at the Centre de Donn{\'e}es astronomiques 
de Strasbourg (CDS) as three separate ascii files. 
A portion is shown for one isotopolog in Table~\ref{supplement} of Appendix~A.

The line lists are quite similar in quantum number coverage among the three minor isotopic 
species and even more so among the two $^{15}$N isotopomers. Therefore, it is not so surprising 
that the parameters determined in the fits are the same and their respective uncertainties are 
quite similar. Some smaller uncertainties of specific H$_2$N$^{13}$CN parameters compared 
to those of the $^{15}$N species are a consequence of lines with higher values of $J$ 
\citep{H2NCN_isos_rot_2011}.
Our initial fits of the old data are summarized in Table~\ref{old-parameters_CA} of 
Appendix~B together with our new parameters for H$_2$N$^{13}$CN. 
The uncertainties of all parameters improved slightly upon addition of our experimental 
data. The main aspect in case of the two $^{15}$N isotopomers is the increase of determined 
spectroscopic parameters from ten to 22 because of the increase in $J$ from 9 to 26 and in 
$K_a$ from 5 to 7. The spectroscopic parameters included in the fit are equivalent to those 
of \citet{H2NCN_isos_rot_2011}; similar to that work, values for $A$, $D_K$, and so on were taken 
from the main isotopic species and the remaining parameters were estimated as described above.

It is also not surprising that the NH$_2$ tunneling splitting is not affected within uncertainties 
by substituting the distant cyano-N atom, only marginally affected by substitution of the closer 
cyano-C atom, and somewhat more by the substitution of the amino-N atom. Replacing one or both 
of the H atoms by D affects the tunneling splitting much more \citep{H2_HD_D2NCN_rot_2013}. 
The heavy atoms are all located close to the $a$-axis, see for example Fig.~3 in 
\citet{H2NCN_isos_rot_2011}, hence heavy atom substitutions do not change the $A$ 
rotational parameter much. In addition, the C atom is quite close to the center of mass 
of the molecule such that its substitution changes $B$ and $C$ only slightly whereas 
replacing one of the N atoms cause larger changes. Changes in higher order spectroscopic 
parameters often reflect the changes in $B + C$ and $B - C$, but exceptions occur, especially 
for some of the higher order parameters.

\section{Astronomical search}
\label{obs_res}

We used the ALMA PILS data to search for the $^{15}$N isotopomers of cyanamide. PILS is 
an unbiased molecular line survey of the Class~0 protostellar binary IRAS 16293$-$2422 
carried out in band 7 of ALMA and covering the spectral range 329.15$-$362.90~GHz at 
0.244~MHz resolution (project-id: 2013.1.00278.S, PI: Jes K. 
J{\o}rgensen\footnote{http://youngstars.nbi.dk/PILS/}). Details of the survey 
have been presented by \citet{PILS_2016}. 
The spectral sensitivity is also very high, 7$-$10~mJy beam$^{-1}$ channel$^{-1}$ or 
4$-$5 mJy beam$^{-1}$ km~s$^{-1}$ such that line confusion is reached in considerable 
parts of the survey. This high sensitivity is very important for searching for less 
abundant molecules or for minor isotopic species of somewhat more abundant molecules. 
Many of our analyses focused on source~B because of its smaller lines widths of 
$\sim$1~km~s$^{-1}$ compared to around 3~km~s$^{-1}$ for source~A. 
Among the most exciting results was the detection of methyl chloride toward both sources 
as the first interstellar organohalogen compound \citep{MeCl_det_2017} besides numerous 
isotopologs containing D or $^{13}$C, many of them for the first time in space 
\citep{DNCO_etc_2016,PILS_2016,formaldehyde_2018,H2NCN_etc_solar-type_2018,nitriles_2018,div-COMs_2018}.


\begin{figure}
\centering
 \includegraphics[width=8cm,angle=0]{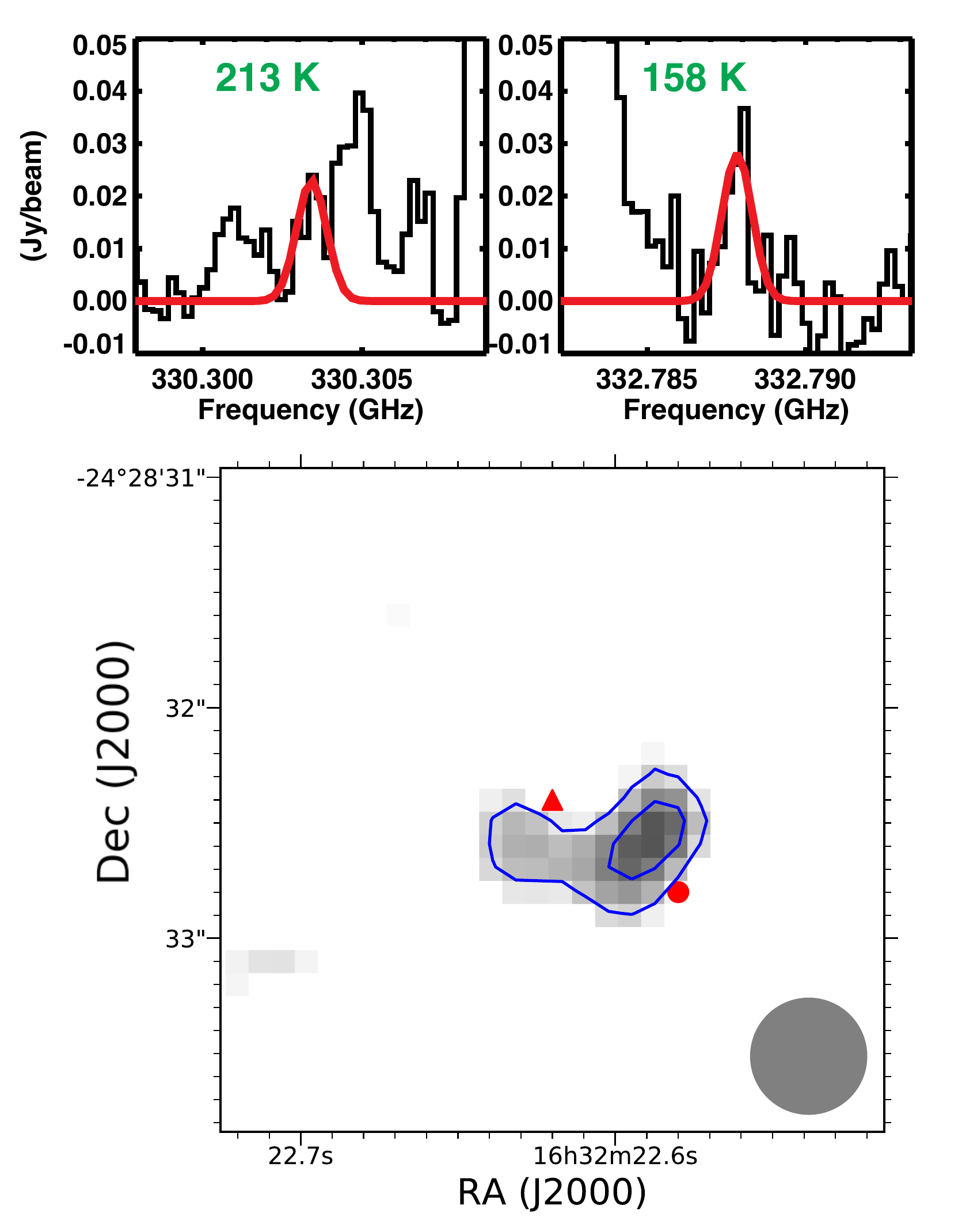}

\caption{Upper panels: Sections of the Protostellar Interferometric Line Survey toward a position 
   one beam offset from source~B displaying two potential lines of H$_2^{15}$NCN. The model 
   described in Section~\ref{obs_res} is indicated in red. The $E_{\rm up}$ values of the two 
   transitions are indicated in green in the upper part of each panel. 
   Lower panel: Integrated intensity map of the transition at 322.7878~GHz. 
   The blue contours correspond to 5 and 7~$\sigma$. The red triangle indicates the 
   position of the continuum peak position, while the red circle indicates the 
   position we analyzed (one beam offset).}
\label{15N-cyanamide_astro}
\end{figure}


\citet{H2NCN_etc_solar-type_2018} analyzed column densities of cyanamide isotopologs toward 
a position offset by 0.5$\arcsec$ from the continuum peak of source~B 
($\alpha_{\rm J2000} =16^{\rm h}32^{\rm m}22\fs58$, $\delta_{\rm J2000} = -24\degr28\arcmin32.8\arcsec$).
This position was chosen to minimize the effects of line self-absorption, which was particularly 
prominent for the main isotopolog of cyanamide. 
The emission lines were modeled under the assumption of local thermodynamic equilibrium, 
which is reasonable at the high densities and high temperatures of source~B, with an excitation 
temperature $T_{\rm ex} = 300$~K, a source size of 0.5$\arcsec$, and  a full line width at 
half maximum (\textit{FWHM}) of 1~km~s$^{-1}$. A column density of $7 \times 10^{13}$~cm$^{-2}$ 
was obtained for the H$_2$N$^{13}$CN isotopolog. Assuming a standard $^{12}$C/$^{13}$C ratio, 
the abundance of the main isotopolog is about $4.8 \times 10^{15}$~cm$^{-2}$.

As a result of our spectroscopic investigation, we can now search for the $^{15}$N 
isotopomers of cyanamide with confidence. The molecules have several lines 
in the range of PILS, but most of them were well below the limit of detection or, 
in some cases, blended with bright lines of other species. 
We find two lines that can potentially be assigned to H$_2^{15}$NCN 
(see Fig.~\ref{15N-cyanamide_astro}) and none for H$_2$NC$^{15}$N. We carefully checked 
that the two potential lines of H$_2^{15}$NCN were not due to any other species already 
identified toward this source and that no line expected to be observed was missing. 
Using the same position, $T_{\rm ex}$, source size, and $FWHM$ as for H$_2$NCN, 
we obtain a good model for a column density of $6 \times 10^{13}$~cm$^{-2}$ 
(see Fig.~\ref{15N-cyanamide_astro}). As the spectra of IRAS 16293$-$2422 are 
very line-rich, more transitions need to be detected to claim a secure detection. 
For now, this column density can only be considered as an upper limit. 
A 3$\sigma$ upper limit of $3 \times 10^{13}$~cm$^{-2}$ is derived for H$_2$NC$^{15}$N. 
This corresponds to $^{14}$N/$^{15}$N ratios of $\gtrsim$80 for H$_2^{15}$NCN and 
$\gtrsim$160 for H$_2$NC$^{15}$N. For comparison in the same source, the $^{14}$N/$^{15}$N 
ratio of CH$_3$CN is about 250 \citep{nitriles_2018}, while a similar lower limit of 
100 was derived for NH$_2$CHO \citep{DNCO_etc_2016}. Single-dish observations of two and 
one transitions of HCN and HNC isotopologs led to ratios of $163 \pm 20$, $190 \pm 38$, 
and $242 \pm 32$, respectively \citep{Wampfler2014}.

More sensitive observations are required to confirm the detection of H$_2^{15}$NCN 
and possibly identify H$_2$NC$^{15}$N in IRAS 16293$-$2422. Such observations may be more 
promising at lower frequencies because line blending and issues due to absorption are 
likely smaller. In addition, observations are, for the most part, easier at lower frequencies. 
On the other hand, the intensities of cyanamide emission drop fast at lower frequencies. 
Therefore, we inspected a synthetic spectrum of the 200$-$320~GHz region for bright lines 
of the $^{15}$N isotopomers of cyanamide that are not blended with lines of species in a 
synthetic spectrum which includes most of the species identified in the PILS data with 
certainty. About 20 lines of H$_2^{15}$NCN and less than 10 lines of H$_2$NC$^{15}$N 
are currently predicted to be free of blending. Since most of the blends in the PILS 
Band~7 data are caused by known species, it may well be that a fair fraction of the about 
30 lines of cyanamide containing $^{15}$N will eventually turn out to be unblended.

\section{Conclusion and outlook}
\label{conclusion}

Our studies of isotopic cyanamide yielded greatly extended experimental line lists of 
H$_2^{15}$NCN and H$_2$NC$^{15}$N up to 507~GHz. The resulting spectroscopic parameters 
should permit sufficiently accurate extrapolation in $J$ up to $\sim$1~THz, that is, 
throughout all presently available ALMA bands. Extrapolation in $K_a$ may be 
questionable because of vibration-rotation interaction, but the covered range should 
be sufficient for all minor isotopic species. Transitions between the tunneling states, 
which obey $c$-type selection rules, cannot be predicted with meaningful accuracy. 
These transitions are quite weak and have, to the best of our knowledge, not been 
identified in space even for the main isotopic species. The line list of H$_2$N$^{13}$CN 
has been extended somewhat in the course of the present investigation, 
and the spectroscopic parameters have been improved.

Calculated line lists of these three isotopic species, including information on their 
intensities and their calculated uncertainties, will be provided or updated in the catalog 
section\footnote{https://cdms.astro.uni-koeln.de/classic/entries/} of the Cologne Database for 
Molecular Spectroscopy, CDMS \citep{CDMS_3}. Files containing the experimental lines or 
parameters along with auxiliary files are available in the data section of the 
CDMS\footnote{https://cdms.astro.uni-koeln.de/classic/predictions/daten/Cyanamide/}.

A first attempt to search for the $^{15}$N isotopologs of cyanamide in the PILS data 
turned out to be negative. The two lines assignable to H$_2^{15}$NCN would imply a low 
$^{14}$N/$^{15}$N ratio if their identity were to be confirmed. Future, more sensitive 
observations of IRAS 16293$-$2422 are required to improve the upper limits or even lead to 
an unambiguous detection of at least one of the isotopic species.


\begin{acknowledgements}
The measurements in K{\"o}ln were supported by the Deutsche Forschungsgemeinschaft (DFG) 
in the framework of the collaborative research grant SFB~956, project B3. 
A.C.'s postdoctoral grant is funded by the ERC Starting Grant 3DICE (grant agreement 336474) 
under the European Union’s Horizon 2020 research and innovation program. 
A.C. is grateful to COST (European Cooperation in Science and Technology) for a STSM grant 
from the COST Action CM1401 ``Our Astro-Chemical History''. O.Z. is funded by the DFG via 
the Ger{\"a}tezentrum ``Cologne Center for Terahertz Spectroscopy''. 
The research of J.K.J. is supported by the European Research Council through the ERC 
Consolidator Grant “S4F” (grant agreement No~646908) and Centre for Star and Planet Formation 
funded by the Danish National Research Foundation.
This paper makes use of the following ALMA data: ADS/JAO.ALMA$\#$2013.1.00278.S. 
ALMA is a partnership of ESO (representing its member states), NSF (USA) and NINS (Japan), 
together with NRC (Canada), NSC  and ASIAA (Taiwan), and KASI (Republic of Korea), in 
cooperation with the Republic of Chile. The Joint ALMA Observatory is operated by ESO, 
AUI/NRAO and NAOJ. Our research benefited from NASA's Astrophysics Data System (ADS).
\end{acknowledgements}


\section*{Appendix A. Supplementary material}
\label{Appendix_A}


\begin{table*}
\begin{center}
\caption{Assigned transitions for H$_2^{15}$NCN, observed transition frequency 
(MHz), experimental uncertainty Unc. (MHz), residual O$-$C between observed 
frequency and that calculated from the final set of spectroscopic parameters, 
weight for blended lines, and source of line.}
\label{supplement}
\begin{tabular}{rrrrrrrrr@{}llrrl}
\hline \hline
$J'$ & $K_a'$ & $K_c'$ & $\varv _t'$ & $J''$ & $K_a''$ & $K_c''$ & $\varv _t''$ & 
\multicolumn{2}{c}{Frequency} & Unc. & O$-$C & Weight \\
\hline
  1 & 0 &  1 & 0 &  0 & 0 &  0 & 0 &   19458&.1     & 0.1     & $-$0.15777 &        & \citet{H2NCN_isos_rot_1972} \\
  2 & 0 &  2 & 0 &  1 & 0 &  1 & 0 &   38916&.282   & 0.100   &    0.00717 &        & \citet{H2NCN_isos_rot_1985} \\
  7 & 0 &  7 & 0 &  6 & 0 &  6 & 0 &  136194&.32870 & 0.05000 &    0.00263 &        & \citet{H2NCN_isos_rot_2011} \\
  8 & 0 &  8 & 0 &  7 & 0 &  7 & 0 &  155645&.83281 & 0.05000 & $-$0.01287 &        & \citet{H2NCN_isos_rot_2011} \\
  9 & 0 &  9 & 0 &  8 & 0 &  8 & 0 &  175095&.43308 & 0.05000 & $-$0.00786 &        & \citet{H2NCN_isos_rot_2011} \\
    &   &    &   &    &   &    &   &        &       &         &            &        &                             \\
 21 & 6 & 15 & 1 & 20 & 6 & 14 & 1 &  407549&.715   & 0.050   &    0.05275 & 0.5000 & Koeln                       \\
 21 & 6 & 16 & 1 & 20 & 6 & 15 & 1 &  407549&.715   & 0.050   &    0.05275 & 0.5000 & Koeln                       \\
 25 & 6 & 19 & 1 & 24 & 6 & 18 & 1 &  485115&.812   & 0.050   &    0.04584 & 0.5000 & Koeln                       \\
 25 & 6 & 20 & 1 & 24 & 6 & 19 & 1 &  485115&.812   & 0.050   &    0.04584 & 0.5000 & Koeln                       \\
 26 & 6 & 20 & 1 & 25 & 6 & 19 & 1 &  504502&.373   & 0.030   & $-$0.03347 & 0.5000 & Koeln                       \\
 26 & 6 & 21 & 1 & 25 & 6 & 20 & 1 &  504502&.373   & 0.030   & $-$0.03347 & 0.5000 & Koeln                       \\
\hline
\end{tabular}
\end{center}
\tablefoot{
This table and those of other isotopologs are available in their 
entirety in the electronic edition in the online journal: 
http://cdsarc.ustrasbg.fr/cgi-bin/VizieR?-source=J/A+A/Vol/Num. 
A portion is shown here for guidance regarding its form and content. 
The quantum number designators $\varv _t = 0$ and 1 indicate the tunneling 
states $0^+$ and $0^-$, respectively.
}
\end{table*}

\section*{Appendix B. Fits of previous data}
\label{Appendix_B}

\begin{table*}
\begin{center}
\caption{Initial spectroscopic parameters $X^a$ (MHz) and their differences $\Delta X^b$ (MHz) of cyanamide isotopologs 
    derived from the previously published data. Our present data are given for H$_2$N$^{13}$CN to allow a direct comparison}
\label{old-parameters_CA}
\renewcommand{\arraystretch}{1.20}
\begin{tabular}[t]{lr@{}lr@{}lr@{}lr@{}l}
\hline \hline
Parameter & \multicolumn{2}{c}{H$_2^{15}$NCN} & \multicolumn{2}{c}{H$_2$NC$^{15}$N} & \multicolumn{2}{c}{H$_2$N$^{13}$CN} & \multicolumn{2}{c}{H$_2$N$^{13}$CN, new} \\
\hline
$\Delta E$                     & 1462540&.~(106)      & 1486091&.~(133)      & 1486783&.~(96)       & 1486522&.~(66)       \\
$F_{ac}$                       &     342&.74~(23)     &     335&.42~(41)     &     345&.999~(69)    &     345&.963~(59)    \\
$F_{ac}^K$                     &    $-$1&.442         &    $-$1&.415         &    $-$1&.457         &    $-$1&.457         \\
$F_{ac}^J \times 10^3$         &       1&.156         &       1&.125         &       1&.2388~(130)  &       1&.2514~(108)  \\
$F_{ac}^{JK} \times 10^6$      &   $-$48&.69          &   $-$47&.52          &   $-$50&.54          &   $-$50&.54          \\
$F_{ac}^{JJ} \times 10^9$      &      10&.64          &      10&.32          &      11&.34          &      11&.34          \\
$A$                            &  308298&.            &  308298&.            &  308209&.3~(182)     &  308205&.1~(122)     \\
$\Delta A/2$                   & $-$3844&.            & $-$3844&.            & $-$3842&.81          & $-$3842&.81          \\
$B$                            &    9845&.1075~(35)   &    9794&.5437~(55)   &   10117&.07913~(141) &   10117&.07868~(93)  \\
$\Delta B/2$                   &    $-$8&.7495~(36)   &    $-$8&.0946~(61)   &    $-$8&.56355~(140) &    $-$8&.56394~(93)  \\
$C$                            &    9604&.1941~(33)   &    9555&.5997~(44)   &    9862&.34845~(140) &    9862&.34933~(114) \\
$\Delta C/2$                   &    $-$0&.28904~(77)  &       0&.2699~(49)   &       0&.35121~(136) &    $-$0&.35051~(110) \\
$D_K$                          &      36&.042         &      36&.042         &      36&.070         &      36&.070         \\
$\Delta D_K/2$                 &    $-$8&.086         &    $-$8&.086         &    $-$8&.092         &    $-$8&.092         \\
$D_{JK} \times 10^3$           &     367&.767~(92)    &     358&.168(~109)   &     376&.911~(169)   &     377&.325~(101)   \\
$\Delta D_{JK}/2 \times 10^3$  &   $-$19&.186~(139)   &   $-$17&.628~(168)   &   $-$17&.972~(102)   &   $-$18&.097~(83)    \\
$D_J \times 10^3$              &       3&.6056~(207)  &       3&.4887~(218)  &       3&.76075~(44)  &       3&.75968~(31)  \\
$\Delta D_J/2 \times 10^3$     &       0&.0165~(204)  &       0&.0405~(221)  &       0&.01861~(34)  &       0&.01813~(26)  \\
$d_1 \times 10^6$              &  $-$119&.81          &  $-$118&.22          &  $-$130&.23~(44)     &  $-$129&.70~(36)     \\
$\Delta d_1/2 \times 10^6$     &       9&.71          &       9&.59          &      10&.54~(44)     &      10&.88~(36)     \\
$d_2 \times 10^6$              &   $-$24&.383         &   $-$23&.984         &   $-$27&.119~(280)   &   $-$27&.739~(219)   \\
$\Delta d_2/2 \times 10^6$     &       4&.733         &       4&.656         &       5&.484~(167)   &       5&.516~(165)   \\
$H_K \times 10^3$              &      10&.76          &      10&.76          &      10&.77          &      10&.77          \\
$\Delta H_K/2 \times 10^3$     &    $-$6&.45          &    $-$6&.45          &    $-$6&.46          &    $-$6&.46          \\
$H_{KJ} \times 10^6$           &  $-$257&.            &  $-$256&.            &  $-$235&.5~(118)     &  $-$196&.2~(45)      \\
$\Delta H_{KJ}/2 \times 10^6$  &      72&.            &      71&.            &      84&.4~(32)      &      82&.3~(7)       \\
$H_{JK} \times 10^6$           &       1&.020         &       1&.012         &       1&.122~(86)    &       0&.945~(44)    \\
$\Delta H_{JK}/2 \times 10^6$  &    $-$0&.273         &    $-$0&.270         &    $-$0&.267~(26)    &    $-$0&.297~(11)    \\
$H_J \times 10^9$              &    $-$0&.761         &    $-$0&.750         &    $-$0&.825         &    $-$0&.825         \\
$\Delta H_J/2 \times 10^9$     &       0&.189         &       0&.186         &       0&.204         &       0&.204         \\
$L_K \times 10^6$              &      16&.0           &      16&.0           &      16&.04          &      16&.04          \\
$\Delta L_K/2 \times 10^6$     &    $-$5&.3           &    $-$5&.3           &    $-$5&.32          &    $-$5&.32          \\
$L_{KKJ} \times 10^6$          &    $-$3&.58          &    $-$3&.56          &    $-$4&.953~(239)   &    $-$5&.800~(54)    \\
$\Delta L_{KKJ}/2 \times 10^6$ &       0&.343         &       0&.343         &       0&.353         &       0&.353         \\
$L_{JK} \times 10^9$           &      11&.8           &      11&.7           &      12&.21~(266)    &      17&.44~(94)     \\
rms                            &       0&.039         &       0&.032         &       0&.042         &       0&.043         \\
$\sigma ^b$                    &       0&.61          &       0&.58          &       0&.87          &       0&.98          \\
\hline
\end{tabular}
\end{center}
\tablefoot{
Watson's $S$ reduction has been used in the representation $I^r$. 
$^{(a)}$ Numbers in parentheses are one standard deviation in units of the least significant figures. 
         Parameters without uncertainties were estimated and kept fixed in the analyses. 
         $X = (X(0^+) + X(0^-))/2$ and $\Delta X  = X(0^-) - X(0^+)$. 
$^{(b)}$ Reduced standard deviation (unitless).
}
\end{table*}


\end{document}